# Phase Transition of single-layer vanadium diselenide on Au(111) with distinguished electronic structures

*Jinbang Hu, Xiansi Wang, Chaoqin Huang, Fei Song and Justin W Wells*

Jinbang Hu

Department of Physics, Norwegian University of Science and Technology, NO-7491 Trondheim, Norway

E-mail: jinbang.hu@ntnu.no

Xiansi Wang

Hunan University, Changsha, 410082, China

Fei Song, Chaoqin Huang

Shanghai Advanced Research Institute, Chinese Academy of Sciences, Shanghai 201000, China;

Justin W. Wells

Department of Physics, Norwegian University of Science and Technology, NO-7491 Trondheim, Norway

Semiconductor Physics, Department of Physics, University of Oslo (UiO), NO-0371 Oslo, Norway

**Abstract**

Herein, we report the reversible structural transition of single-layer $VSe_2$ grown on Au(111) through alternating thermal annealing and Se replenishment. Using scanning tunneling microscopy (STM) and angle-resolved photoemission spectroscopy (ARPES), we demonstrate the epitaxial growth of high-quality $VSe_2$ on Au(111) with the octahedral (1T) structure and the Se-vacancy-induced transformation of $VSe_2$ from the metallic moiré (1T) phase to the semiconducting (2H) phase. With convincing agreement between the experimental results and DFT calculations, the nanostructure near the grain boundary in the defective intermediate phase is confirmed, as well as the reaction pathway with Se gradually depleting at elevated temperatures. Importantly, it is revealed that the density of the linear Se defects plays a crucial role in the formation of the 2H domain phase due to the increment of the in-plane lattice parameter after Se desorption and the better thermal stability of the 2H phase compared to the 1T phase. The proper control of the density of Se atoms in the topmost Se layer of $VSe_2$ could feasibly manipulate the ratio between the 1T phase and the 2H phase in the steak-shaped domain, which is regarded as a good platform for 2D homojunctions in nanoelectronics.

## 1. Introduction

Recent advances in artificial two-dimensional (2D) van der Waals structures show great promise for the development of novel electronic and optoelectronic devices[1]. To date, considerable efforts have been devoted not only to investigating fundamental properties but also to exploring engineering strategies to modify local phase transitions in 2D materials, aiming for the controllable design of homojunctions[2]. Atomically thin 2D materials such as transition metal dichalcogenides (TMDs)[3], graphene[4] and graphene analogues[5] have been credited with unique properties, including valley-polarized exciton dynamics[6], 2D ferromagnetism[7], and room-temperature ferroelectricity[8]. In the post-Moore era, a variety of strategies such as vapor-phase deposition[9], laser irradiation[2a, 10], chemical doping[11], and doping engineering[12] have been adopted for realizing the controllable construction of 2D homojunctions.

Apart from the heterojunctions constructed from two different materials, TMDs with naturally matched chemical and electronic structures in different polymorphic phases (e.g., 1T, 2H) provide a convenient platform for fabricating 2D homojunctions, which exhibit continuous band bending and high-quality carrier transport and separation at the interface[2b]. Specifically, phase-engineered metallic 1T-$MoS_2$/semiconducting 2H-$MoS_2$ [13] and metallic 1T′-$MoTe_2$/semiconducting 2H-$MoTe_2$[2a] have been reported to have low-resistance contacts and improved carrier mobility. Therefore, corroborating the actual dynamical process of the transformation between 1T and 2H phases and their boundary structures is crucial for constructing 2D homojunctions.

Here, we report the synthesis of epitaxial single-layer (SL) $VSe_2$ on Au(111) and provide in-situ observations of the phase transformation process from the pristine homogeneous 1T phase, through a defective steak-shaped domain phase with the coexistence of the 1T phase and the 2H phase, to a VSe striped phase decorated by lines of Se atoms upon annealing at elevated temperatures. Our study employs a diverse set of characterization methods, combining STM with ARPES to highlight the differences between the 1T phase and the 2H phase. Throughout the exposition, we use density functional theory (DFT) calculations to explain the observations. We discuss the structural defect near the grain boundary and reaction pathway determined by the loss of Se in the depleted structure with an increased in-plane lattice constant, which will facilitate the precise design of artificial 2D van der Waals structures as desired homojunctions in the future.

## 2. Experiment details and calculation methods

All sample preparation steps and experiments were performed under ultrahigh vacuum conditions better than $4\times10^{-10}$ mbar. The Au(111) surface was prepared by repeated sputtering and annealing cycles. The quality of the clean surface was confirmed by the well-known Au-herringbone reconstruction in STM observation and the Shockley state in ARPES measurements (see Figure S1 in the Supplemental Material). Subsequently, Se (purity 99.9999%) and V (purity 99.99%) were co-deposited onto the clean surface at 300 °C, yielding a pure 2D moiré structure approximately at the monolayer. The prepared SL $VSe_2$ was then annealed at elevated temperatures up to 460 °C. The different structures formed after the sample preparation were confirmed by LEED and STM, as well as corresponding band structure measurements. The lab-based ARPES measurements were performed at room temperature using an aberration-corrected, energy-filtered photoemission electron microscope (EF-PEEM) (NanoESCA III, Scienta Omicron GmbH) equipped with a focused helium discharge lamp primarily generating He I photons at $h\nu$ = 21.22 eV. A pass energy ($E_P$) of 25 eV and a 0.5 mm entrance slit to the energy filter were used[14], yielding nominal energy and momentum resolutions of $\Delta E$ = 50 meV and $\Delta k$=0.02 Å$^{-1}$.

The density functional theory (DFT) calculations were performed using the Vienna Ab initio Simulation Package (VASP)[15]. The interactions between the valence electrons and ion cores were described by the projector augmented wave method[16]. The electron exchange and correlation energy was treated using the generalized gradient approximation with the Perdew-Burke-Ernzerhof functional[15, 16b]. The kinetic energy cutoff of the plane-wave basis was set to 500 eV by default. A 5×5×1 supercell of the 1T phase on a Au(111) substrate was considered to study the effect of the moiré superstructure. The STM simulations were performed using the Tersoff–Hamann approach[17]. The band structures for the freestanding SL 1T, 2H and VSe buckling phases were calculated without spin-orbital coupling (SOC). The calculated bands of the SL 2H phase and the VSe striped phase are shifted down by 1.0 eV and 4.3 eV, respectively, to match the ARPES data. All the structures were optimized until the forces on the atoms were less than 20 meV Å$^{-1}$. The vacuum layer was thicker than 15 Å in order to eliminate spurious interaction between two adjacent slabs.

## 3. Discussion

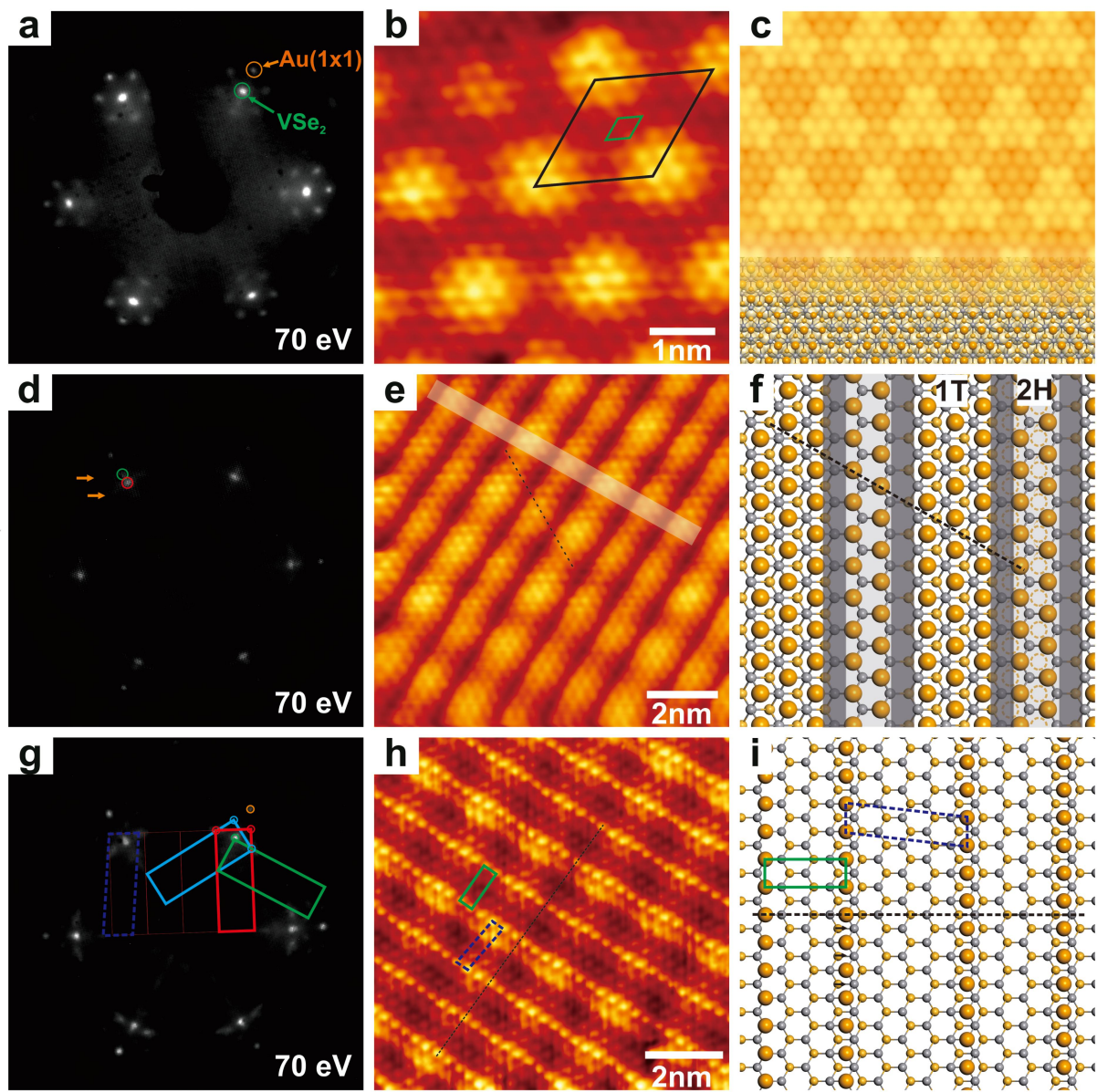

Figure 1. Evolution of the crystal morphology of SL VSe$_2$ on Au(111) at elevated temperatures. Top row shows (a) LEED pattern (70 eV), (b) STM observation and corresponding energetically favourable atomic model (c) of an epitaxially grown VSe$_2$ monolayer at 300 ℃, the STM simulation is overlaid on the top of the image in panel (c). The middle row and the bottom row show further annealing to 380 ℃ (d-f) and 460 ℃ (g-h), respectively. The circles in (b) show the positions of LEED spots for VSe$_2$ monlayer before (green circle) and after annealing to 380 ℃ (yellow circle). The corresponding unit cells of the selenide linear reconstruction in (g-h) are indicated by the rectangular and dashed parallelogram. The black dashed lines in (e,f,h,i) are used to mark the relative position of topmost Se atoms. Scanning parameters: (b) 1 V, 1.0 nA; (e) -1.2V, 6.9 nA; (h) -1.2V, 8.4 nA;

Figure 1 depicts a general overview of three different phases obtained in epitaxially grown SL VSe$_2$ at 300 ℃, and subsequently annealed to 380 ℃ and 460 ℃, as demonstrated by LEED patterns and STM images, respectively. Initially, an SL VSe$_2$ was fabricated by Se and V co-deposited onto the clean Au(111) at 300 ℃. From the LEED panel in Figure 1a, a clear moiré pattern surrounding the main spots of the VSe$_2$ atomic lattice, which coincides with the Au(1x1) diffraction spots, indicates a possible commensuration between a (6×6) reconstruction of the Au(111) substrate and the apparent moiré corrugation with a (5×5) superlattice of the as-grown VSe$_2$, which is in agreement with the STM observation shown in Figure 1b. Based on the STM data, the atomic lattice constant of the moiré superstructure has a lattice parameter of 17.7 Å, while the atomic lattice parameter is 3.45 Å, as illustrated by the unit cells marked by the black and green rhombuses in Figure 1b, respectively. To further investigate the atomic structure of VSe$_2$ on Au(111), we employed DFT calculations to construct an optimized 5×5 structural model of the 1T-VSe$_2$ on the Au substrate, as depicted in Figure 1c. The corresponding simulated STM image shows a periodic moiré structure, consistent with the STM data. The model of 1T-VSe$_2$ used here is based on our ARPES measurements, which will be discussed below.

Annealing the samples to 380 °C (middle row of Figure 1) induces a phase transition. As evidenced by the LEED pattern (Figure 1d), this pattern shows an additional hexagonal lattice (marked by a red circle) with a slightly smaller reciprocal lattice constant than that of the as-grown 1T phase (marked by a green circle), and the corresponding moiré spots (marked by yellow arrows) mostly disappear. Meanwhile, the STM image (Figure 1e) reveals a different appearance in the moiré structure (marked by the white ribbon) and clearly visible linear defects decorated in the SL VSe$_2$. The density of the linear defects relates to the annealing temperature, as discussed in Ref[18]. The moiré structure in dislocation-dominated defective VSe$_2$ appears without a commensurate relationship to the linear defects. We deduce that this new moiré structure originates from the interaction with the Au-herringbone reconstruction, which can be clearly observed in Figure 2a of ref[18]. Moreover, the topmost Se atoms seem to be rearranged in two adjacent steak-shaped domains, as indicated by the black dashed line in Figure 1e. The dashed line passes through the top site of Se atoms in one steak-shaped domain while going through the bridge site of Se atoms in the other. We propose a possible structural model (Figure 1f) with 1T and 2H phases coexisting in the defective VSe$_2$ monolayer, with the linear Se defects marked by the grey ribbon separating the two different phases. This is similar to what has been previously observed in Se vacancies rearranging to mediate the transformation from 1T domains to 2H domains[9b, 19]. Further confirmation of the existence of the 2H phase will be given in the discussion of ARPES data and STM observation below.

The complete transition to a striped phase is observed after annealing to 460 °C and is shown in the bottom row of Figure 1. The atomically resolved STM image (Figure 1h) reveals the linear rearrangement of the Se atoms with a rectangular unit cell for two nearest bright lines of atoms and a parallelogram unit cell for two bright lines with a shift of half a unit cell along the line of Se atoms. The experimentally observed shift can be clearly identified by the black dashed guideline in the STM image (Figure 1h). Correspondingly, the LEED data of the striped phase in Figure 1g exhibits a complicated diffraction pattern similar to what was observed previously in similarly prepared SL VS$_2$ on Au(111)[20]. Taking into account the unit cell of the structure identified by STM, the diffraction spots (marked by colorful circles) in the LEED pattern can be reproduced by the reciprocal rectangular unit cells derived from the three rotational domains. Moreover, the expected LEED spots from the reconstruction of Se atoms appear as a bright line and are broadened along the short side of the rectangular unit cells, which can be explained by the parallelogram unit cell for two bright Se lines with the shift of half a unit cell, marked as a dashed parallelogram in both Figure 1g and 1h, respectively. Based on the analysis of the LEED data and STM measurements, an optimized structural model of VSe with the line of Se atoms decorated on top as a defective phase of VSe$_2$ was constructed (Figure 1i). Notably, the topmost Se atoms in the line-shaped array can stack on the top site or hollow site with regard to the bottom layer of Se atoms. The parallelogram unit cell for two bright Se lines with the shift of half a unit cell, revealed in the

STM observation, indicates that the corresponding two adjacent lines of Se atoms have the same stacking style. In contrast, the requirement on the ratio between the two sides of the rectangular unit cell for two closer Se lines indicates that one of the line-shaped Se arrays stacks like the 1T phase and another line of Se atoms stacks like the 2H phase. Thus, the two different stacking styles of the Se lines revealed here also indicate the possible phase transition from the 1T phase to the 2H phase in the local steak-shaped domain separated by the linear Se defects. Moreover, compared to the LEED pattern of the as-grown SL VSe$_2$ with a periodic moiré structure, the slightly shrunk reciprocal rectangular unit cell indicates that the VSe with lines of Se atoms stacked on top has an expanded in-plane lattice constant after an obvious Se depletion caused by heating the VSe$_2$ layer in vacuum.

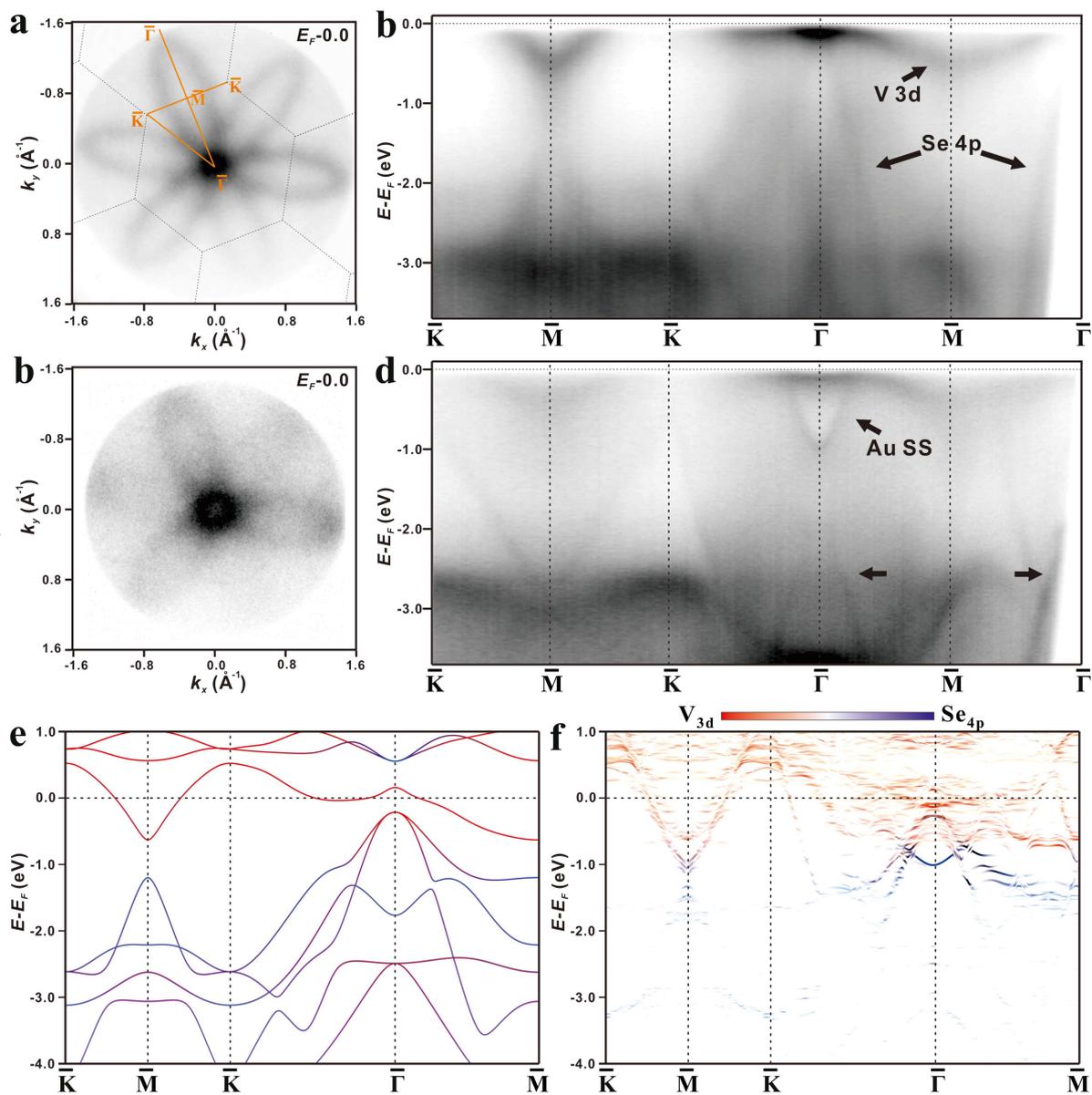

Figure 2. Electronic structure of the as-grown SL VSe$_2$ on Au(111) and upon heating to 380 °C. Fermi surface (a) and band dispersion (b) along the high-symmetry directions of the surface Brillouin zone (SBZ) of the as-grown SL VSe$_2$ measured (hν = 21.22 eV) at room temperature. (c) and (d) Same as (a) and (b), but for the steak-shaped VSe$_2$ (heating to 380 °C) decorated by lines of Se defects. The dashed hexagon in (a) represents the 2D SBZ with symmetry point labels overlaid. A few observations of particular interest are indicated by arrows. The bottom row of the figure shows the nonmagnetic calculated band structure of the SL 1T-VSe$_2$ without (e) and with (f) Au substrate-induced moiré superstructure considered for the model. The different atomic orbital contributions are color coded.

ARPES measurements have been carried out to track the electronic structure of the several phases observed in the STM images upon heating. Figures 2a and 2b display the Fermi surface and band structures along the high symmetry pathway (yellow line) of as-grown SL VSe$_2$, respectively. The new bands indicated by black arrows in Figure 2b are periodic with respect to the SBZ of SL VSe$_2$ and align with the reported band structures of SL 1T-VSe$_2$ grown on an HOPG substrate[21], as well as with nonmagnetic calculations for SL 1T-VSe$_2$, shown in Figure 2e. The orbital-projected band structures of the V 3d and Se 4p orbitals reveal that the Fermi surface in Figure 2a, forming a closed flower-like contour, is dominated by the V 3d-derived band crossing Fermi level ($E_F$) as an electron-like conduction band, while a parabolically dispersing band in Figure 2b, steeply descending from the Γ point into the lower valence band, exhibits primary contributions from Se 4p orbitals. Additionally, the band structures from SL VSe2 displays a fuzzy feature. We infer it may result from the periodic moiré structure of SL VSe$_2$ due to the interaction with the Au substrate. DFT calculations, neglecting SOC effects, were performed using a 5x5 supercell of SL VSe$_2$ on an Au substrate with a periodic moiré structure, inferred from LEED and STM data. Figure 2f shows the calculated band structure with the contributions from V 3d and Se 4p orbitals. The projected band structure near $E_F$ displays a dispersion similar to that of free-standing SL VSe$_2$ but shows a broadening character, indicating that the moiré structure induces distortions in SL VSe$_2$, thereby partially contributing to the fuzzy feature.

Upon heating to 380 °C, Figures 2c and 2d show the Fermi surface and band dispersion of the steak-shaped phase decorated with linear defects (as shown in Figure 1e). Compared to the ARPES data recorded for the as-grown homogeneous 1T phase, the band structure originating from SL 1T-VSe$_2$ weakens and several new band features appear, marked by black arrows in Figure 2d. The weak band feature near $E_F$, similar to the as-grown 1T phase, indicates there are still domains of the 1T phase remaining in the defective SL VSe$_2$. However, a new parabolic band appears with the whole feature at 2 eV below $E_F$, which is extremely different from the parabolic Se 4p-derived band in the 1T phase that disperses with its maximum near $E_F$. Our DFT calculations for the 1T phase, with an increased in-plane lattice constant observed in the defective steak-shaped phase, indicate that the energy position of the parabolically dispersing band (marked by black arrows), mainly from Se 4p orbitals, remains unchanged compared to the band structure of the homogeneous 1T phase (Figure S2). Based on the new parabolic band appearing deeply at 2 eV below $E_F$, while the V 3d-derived band of defective 1T-VSe$_2$ domain remains crossing $E_F$ with minimal energy shift caused by charge transfer from the Au substrate, we deduce that this new parabolic band feature originates from a new phase formed upon annealing. In addition, a parabolic electronic pocket appears around the Γ point with its bottom at -1.0 eV, which is a typical band feature of the Shockley state from the clean Au(111). The Shockley state observed in the ARPES measurements indicates the moiré structure of the dislocation-dominated defective VSe$_2$ recorded in STM measurements (Figure 1e) originates from the interaction with the typical herringbone reconstruction of the Au(111) surface.

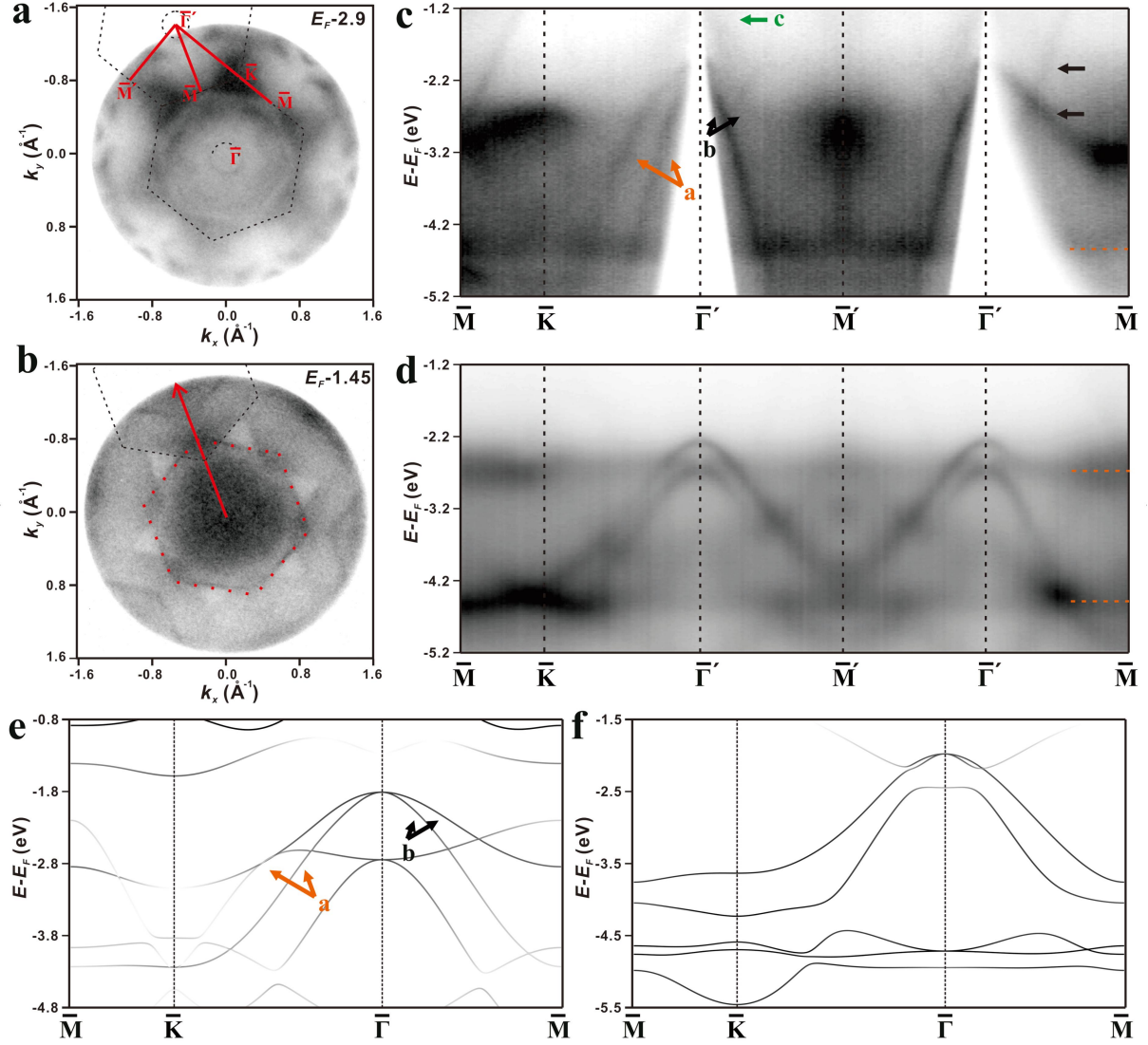

Figure 3. Electronic structure of defective SL VSe$_2$ upon heating to 450 °C (a-c) and 460 °C (d). (a) Constant-energy contour at -2.9 eV with 2D SBZ of defective SL VSe$_2$ shown by a black dashed hexagon. (b) Constant-energy contour at -1.45 eV shows the Umklapp scattering process of the Au sp band. The original (scattering) Au sp band is marked by a red (black) dashed hexagon. The reciprocal lattice vectors marked by a red arrow indicate one of the possible Umklapp scattering process. (c) Band structure recorded on the high symmetry directions marked by the red line in (a). (d) Same as (c), but for VSe (heated to 460 °C) decorated by lines of Se atoms on top. The position of flat bands in (c) and (d) are indicated by a yellow dashed line. The V 3d projected band structure of the SL 2H-VSe$_2$ (e) and SL VSe (f) without SOC included in the calculation are shown as a comparison with experimentally recorded band structures. The new parabolic band structure appearing deeply at 2 eV below E$_F$ are pointed by yellow and black arrows in (b) and (e).

To determine the lattice structure of the new phase consisting of V and Se associated with the parabolic band structure appearing deeply at 2 eV below E$_F$, we further studied the electronic structure of the steak-shaped VSe$_2$ within an energy window near this parabolic band. Figures 3a and 3b show the constant-energy contours at -2.9 eV and -1.45 eV below E$_F$. The band feature shows sixfold symmetry, which indicates the new phase in the domains of steak-shaped VSe$_2$ has a hexagonal lattice structure. Specifically, in Figure 3b, a careful analysis of the hexagon-shaped Au sp band, marked by the dashed line, reveals the Umklapp scattering process of the Au sp band with the reciprocal lattice vectors (red arrow) of the steak-shaped phase. The presence of only one type of reciprocal lattice vector with sixfold symmetry suggests that the new phase in the domains of steak-shaped VSe$_2$ has the same hexagonal lattice parameters as the 1T-VSe$_2$. Hence, a possible candidate, 2H-VSe$_2$, has been considered. This new phase formed upon annealing is similar to the phase transition reported in MoS$_2$, MoSe$_2$, MoTe$_2$ and PtSe$_2$[2a, 9b, 19, 22]. Moreover, referring to the sp band feature from the clean Au(111) substrate (Figure S1b), the hexagonal unit cell with an increased in-plane lattice vector of the proposed candidate 2H-VSe$_2$ and the already confirmed 1T-VSe$_2$ in the steak-shaped domain is calculated to be 3.65 Å, which is in agreement with an increased lattice parameter confirmed from LEED and STM data.

A comparative study between the electronic structure of the steak-shaped VSe$_2$ from the ARPES measurements and the calculated band structure of the SL 2H-VSe2 can be used to identify if the proposed 2H phase was formed in the different domains of the steak-shaped VSe$_2$ upon heating. Figure 3c shows the band dispersion along the high symmetry directions, marked by the red line in Figure 3a. The parabolic bands with two branches along the Γ- M and Γ- K directions, labelled as 'a' and 'b', disperse separately away from the Γ point. The calculated bands of SL 2H-VSe$_2$ without SOC, as displayed in Figure 3e, exhibit overall good agreement with the experimental bands 'a' and 'b' in Figure 3c. As discussed above, the typical V 3d-derived band losing some of its spectral weight but remaining crossed $E_F$ (Figure 2d) confirms the existence of 1T phase in the steak-shaped domain. The new clear band structures at 2 eV below $E_F$ can be explained by our DFT predicted band structures of the 2H phase with a slightly increased lattice instead of assuming the 1T phase. We then conclude that a phase transition from the as-grown homogeneous 1T phase to the coexistence of 1T and 2H domains randomly distributed in the defective steak-shaped phase occurs upon heating. Note that the experimentally recorded band structure of the 1T phase shows minimal energy shift compared to the calculated freestanding SL 1T phase (Figures 2b and 2e). In contrast, the calculated band structure of the freestanding SL 2H phase (Figure 3e) is shifted down by about 1.0 eV to align with the ARPES data. This shift may result from electron doping of the 2H domains due to charge transfer from the substrate, since the 1T phase of VSe$_2$ has been reported to be metallic whereas the 2H phase appears as a semiconductor[23].

Annealing in UHV at higher temperatures of 460 °C leads to a phase transition from the steak-shaped VSe$_2$ to a striped phase in our STM study. We speculate that this striped phase might be a buckling VSe network decorated by lines of Se atoms, which results from losing almost all of the Se atoms in the top layer of VSe$_2$ at higher temperatures. Correspondingly, in Figure 3d, the experimentally recorded band dispersion sliced along the same symmetry directions as Figure 3c has been applied to study additional information about the proposed VSe network. Two similar parabolic bands disperse to a maximum energy at the Γ point with a direct gap of 0.45 eV, which is distinctly different from the band dispersion of the steak-shaped phase in Figure 3c. A nonmagnetic calculation of the simplified VSe network (Figure 3f) ignoring the linear reconstruction from the topmost Se atoms reveals a band dispersion quite similar to our ARPES measurements, except that there are non-dispersive bands (marked by yellow dotted lines) at -2.6 eV, coinciding with the maximum of the lower parabolic band, and another one at -4.5 eV. Since these nondispersive bands remain after a phase transition from the steak-shaped phase to the striped phase, we assume these flat bands may originate from surface resonance states that mix with bulk bands (or bulk band projections)[24]. The overall agreement between experimental and calculated band dispersions, as well as our analysis of the model based on STM observation, indicates a buckling VSe network was formed with the significant depletion of almost all Se atoms in the top layer of VSe$_2$ upon heating. Compared to the initially prepared 1T-VSe$_2$ with the moiré superstructure, we experimentally find an increasing tendency of the lattice parameter during the structure transformation in our ARPES measurements, which are consistent with the observations from

LEED and STM. Note that the overall band structure of the VSe striped structure rigidly shifts in our ARPES measurements compared to the calculation of the freestanding monolayer VSe without the Au substrate and doping of Se arrays considered, similar to what has been observed in hBN on Ir(111)[25].

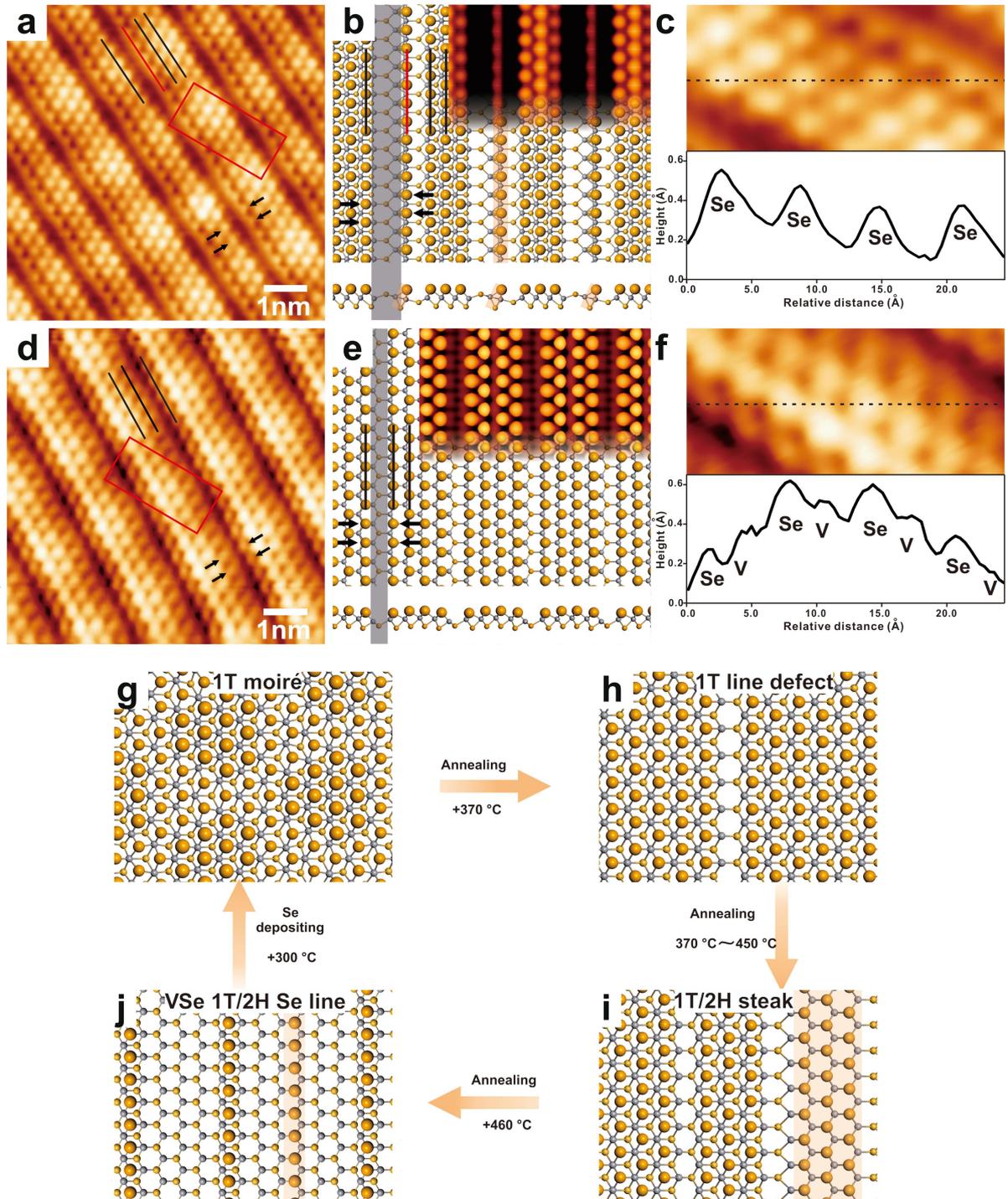

Figure 4. Atomically resolved STM of two different steak-shaped pattern (a,d) upon heating and schematic illustration of the structural transformation from pure 1T $VSe_2$ to a VSe striped phase (g-j). (a) STM measurement of the 1T steak-shaped phase decorated by linear Se defects, (b) The proposed energetically favourable atomic model with simulated STM image overlaid to the top right side of the model. (c-d) Same as (a-b) but for the 2H steak-shaped phase. The relative Se atomic positions in different steak-shaped domains near the sides of the linear defect are marked by black arrows in (a,b,d,e), respectively. The short lines in the STM image (a,d) are used to mark the relative position of the Se atoms near the linear defect, which is marked by the gray shadow in (b,e). (c) and (f) Enlarged images of the region indicated by the red rectangle in images (a) and (d). The intensity profile along the black dashed line is shown below the corresponding image. Scanning parameters: (a) -1.2 V, 6.9 nA; (d) -1.2 V, 1.0 nA.

On the basis of the above experimental data, STM and ARPES measurements reveal three distinct crystalline phases. Specifically, the as-grown 1T-VSe$_2$ with periodic moiré structure and the complete transition to the VSe stripe phase decorated with lines of Se atoms have been clearly discussed. However, the intermediate phase, the steak-shaped VSe$_2$, confirmed to be a coexistence of 1T and 2H domains from our analysis of ARPES measurements and DFT calculations, remains unclear, especially the structure near the linear defects, for which a complete understanding is important in the scenario of artificially designing 2D van der Waals structures. Here, atomically resolved STM images reveal two different steak-shaped structures in Figures 4a and 4d, together with the optimized configurations in Figures 4b and 4e. We refer to these two different steak-shaped phases as 'phase I and II' shown in Figure 4a and 4d, respectively. At first glance, the domain of phase I exhibits protrusions that are bright on one edge but faint on the other, indicating the linear defective structures near the two sides of the steak-shaped domain in phase I are different. Phase II, by contrast, appears brighter in the middle than at either of the two edges. Moreover, a careful analysis of the relative Se position within the steak-shaped domain (indicated by the black arrows in STM image) shows that the Se atoms on one side of the linear defect shift by half a unit cell along the direction of the line relative to the Se atoms on the other side in phase I, whereas there are no obvious shifts of the Se atoms on either side of the linear defect in phase II. Combined with the structural characteristic that the linear defect in phase II appears narrower than in phase I, as indicated by the short lines near the top of the STM images (Figures 4a and 4d), we deduce that the adjacent steak-shaped domains in phase I are separated by one line of Se defects, while the adjacent domains in phase II are separated by two adjacent lines of Se defects.

Interestingly, comparing the two enlarged steak-shaped domain phases in Figure 4c and 4f, the intensity profile across the domain in the upper level of the corresponding figure shows an obvious difference. Alternate Se-V peaks are observed in the domain of phase I, while only Se peaks are observed in phase II, resembling the as-grown 1T phase (Figure S3a-b) and what has been seen previously in similarly prepared SL MoS$_2$, MoSe$_2$ and PtSe$_2$[2a, 9b, 19, 22]. Moreover, the consistent hexagonal and honeycomb features are also confirmed in our STM simulations of the freestanding SL 1T and 2H phases, respectively (Figure S3c-d). Combined with our ARPES confirmation of the coexistence of the 1T phase and the 2H phase, we deduce that the steak-shaped domain with linear Se defects in phase I is the 1T phase, and phase II is the 2H phase. Based on our discussion of the relative position of Se atoms near the linear defect, the corresponding energetically favourable atomic models are proposed in Figure 4b and 4e. Simulations of the STM images (top-right of Figures 4d and 4e) based on the proposed model show overall good agreement with the atomic resolution STM images (Figure 4a and 4d). Specifically, a bamboo-shaped feature of the line of Se atoms on one side of the steak-shaped domain in phase I (marked by the red line in Figure 4a) is well reproduced in the simulated image (top-right of Figure 4b). A likely explanation for this abnormal distorted feature at the boundary of one side of the steak-shaped domain is that this array of Se atoms is at a transition state between the 1T phase and the 2H phase, due to a higher electronic contribution from V atoms in the 2H phase distributed between two adjacent Se atoms.

The atomic mechanism of the phase transition from the homogeneous 1T phase to the striped phase of VSe with linear Se atoms decorated on top is presented in Figure 4g-i. Specifically, thermal annealing of the as-grown 1T VSe$_2$ film at an elevated temperature leads to the loss of top-layer Se atoms, forming a defective structure accompanied by an increased lattice constant. The increasing desorption of the Se atoms from the top layer significantly impacts the rearrangement of the remaining atoms in different domains. The local steak-shaped domain separated by the linear Se defects retains the 1T phase when a small amount of Se desorption occurs, as the lattice constant rarely changes. Furthermore, the remaining Se atoms to form the domain in the 2H phase occurs after increased Se desorption, indicating that the 2H phase is more stable in a slightly larger lattice unit cell than the unit cell for the homogeneous 1T phase. Heating to 460 °C leads to a complete phase transformation from the steak-shaped VSe$_2$ to a striped VSe with a small amount of Se decorated on top (Figure 4i), and this structure can, of course, be transformed back into pristine defect-free 1T phase by adding Se atoms while annealing the sample to 300 °C in vacuum[18]. Based on the discussion of the structural transformation, we deduce that the density of the linear Se defects plays an important role in the formation of the 2H domain phase. Hence, proper control of the density of Se atoms by thermal annealing and re-adding Se atoms can adjust the ratio between the 1T phase and the 2H phase in the steak-shaped domain.

## 4. Conclusion

In conclusion, we have reported a phase transition from the initial homogeneous VSe$_2$ 1T phase fabricated on Au(111), passing through defective 1T/2H coexisting intermediates, to the formation of the final striped VSe with a linear reconstruction of Se atoms on top. Characterizations by STM, ARPES, and their agreement with DFT calculations elucidate the difference in the atomic and electronic structures between the 1T phase and the 2H phase. Our clarification of the defective structure near the boundary between two VSe$_2$ domains reveals the dynamics of defect movement and the rearrangement behavior of the remaining Se atoms. We find that the desorption of Se atoms leads to a defective lattice formed with increased lattice parameters, which is more stable for the 2H phase at high temperatures. Moreover, the reversible phase transition pathway indicates that proper control of the density of Se atoms by thermal annealing and re-adding Se atoms can be regarded as a good method for artificially creating 2D van der Waals structures as homojunctions.

**Supporting Information**

Supporting Information is available from the Wiley Online Library or from the author.

**Acknowledgements**

This work was partly supported by the Research Council of Norway, project numbers 262 633, 324 183, 315 330, and 335 022. The STM measurement was financially supported by the National Key Research and Development Program of China (2021YFA1600800) and National Natural Science Foundation of China (11874380, 22002183). The DFT calculations in collaboration with X.S. Wang supported from the National Natural Science Foundation of China (Grants No. 11804045 and 12174093) and the Fundamental Research Funds for the Central Universities.

# Reference

[1] a) Q. H. Wang, K. Kalantar-Zadeh, A. Kis, J. N. Coleman, M. S. Strano, *Nature Nanotechnology* **2012**, 7, 699; b) K. S. Novoselov, A. Mishchenko, A. Carvalho, A. H. Castro Neto, *Science* **2016**, 353, aac9439.

[2] a) S. Cho, S. Kim, J. H. Kim, J. Zhao, J. Seok, D. H. Keum, J. Baik, D.-H. Choe, K. J. Chang, K. Suenaga, S. W. Kim, Y. H. Lee, H. Yang, *Science* **2015**, 349, 625; b) F. Wang, K. Pei, Y. Li, H. Li, T. Zhai, *Advanced Materials* **2021**, 33, 2005303.

[3] S. Manzeli, D. Ovchinnikov, D. Pasquier, O. V. Yazyev, A. Kis, *Nature Reviews Materials* **2017**, 2, 17033.

[4] a) K. S. Novoselov, V. I. Fal'ko, L. Colombo, P. R. Gellert, M. G. Schwab, K. Kim, *Nature* **2012**, 490, 192; b) A. K. Geim, K. S. Novoselov, *Nature Materials* **2007**, 6, 183.

[5] F. Reis, G. Li, L. Dudy, M. Bauernfeind, S. Glass, W. Hanke, R. Thomale, J. Schäfer, R. Claessen, *Science* **2017**, 357, 287.

[6] P. Rivera, K. L. Seyler, H. Yu, J. R. Schaibley, J. Yan, D. G. Mandrus, W. Yao, X. Xu, *Science* **2016**, 351, 688.

[7] a) D. J. O'Hara, T. Zhu, A. H. Trout, A. S. Ahmed, Y. K. Luo, C. H. Lee, M. R. Brenner, S. Rajan, J. A. Gupta, D. W. McComb, R. K. Kawakami, *Nano Letters* **2018**, 18, 3125; b) Z. Zhang, J. Shang, C. Jiang, A. Rasmita, W. Gao, T. Yu, *Nano Letters* **2019**, 19, 3138.

[8] a) S. Yuan, X. Luo, H. L. Chan, C. Xiao, Y. Dai, M. Xie, J. Hao, *Nature Communications* **2019**, 10, 1775; b) K. Chang, J. Liu, H. Lin, N. Wang, K. Zhao, A. Zhang, F. Jin, Y. Zhong, X. Hu, W. Duan, Q. Zhang, L. Fu, Q.-K. Xue, X. Chen, S.-H. Ji, *Science* **2016**, 353, 274.

[9] a) J. H. Sung, H. Heo, S. Si, Y. H. Kim, H. R. Noh, K. Song, J. Kim, C.-S. Lee, S.-Y. Seo, D.-H. Kim, H. K. Kim, H. W. Yeom, T.-H. Kim, S.-Y. Choi, J. S. Kim, M.-H. Jo, *Nature Nanotechnology* **2017**, 12, 1064; b) X. Lin, J. C. Lu, Y. Shao, Y. Y. Zhang, X. Wu, J. B. Pan, L. Gao, S. Y. Zhu, K. Qian, Y. F. Zhang, D. L. Bao, L. F. Li, Y. Q. Wang, Z. L. Liu, J. T. Sun, T. Lei, C. Liu, J. O. Wang, K. Ibrahim, D. N. Leonard, W. Zhou, H. M. Guo, Y. L. Wang, S. X. Du, S. T. Pantelides, H. J. Gao, *Nature Materials* **2017**, 16, 717.

[10] Y. Yu, M. Ran, S. Zhou, R. Wang, F. Zhou, H. Li, L. Gan, M. Zhu, T. Zhai, *Advanced Functional Materials* **2019**, 29, 1901012.

[11] a) G. Eda, H. Yamaguchi, D. Voiry, T. Fujita, M. Chen, M. Chhowalla, *Nano Lett* **2011**, 11, 5111; b) J. Wu, J. Peng, Y. Zhou, Y. Lin, X. Wen, J. Wu, Y. Zhao, Y. Guo, C. Wu, Y. Xie, *Journal of the American Chemical Society* **2019**, 141, 592.

[12] a) D. Xiang, T. Liu, J. Xu, J. Y. Tan, Z. Hu, B. Lei, Y. Zheng, J. Wu, A. H. C. Neto, L. Liu, W. Chen, *Nature Communications* **2018**, 9, 2966; b) N. Liu, H. Tian, G. Schwartz, J. B. H. Tok, T.-L. Ren, Z. Bao, *Nano Letters* **2014**, 14, 3702.

[13] R. Kappera, D. Voiry, S. E. Yalcin, B. Branch, G. Gupta, A. D. Mohite, M. Chhowalla, *Nature Materials* **2014**, 13, 1128.

[14] M. Escher, N. Weber, M. Merkel, B. Krömker, D. Funnemann, S. Schmidt, F. Reinert, F. Forster, S. Hüfner, P. Bernhard, C. Ziethen, H. J. Elmers, G. Schönhense, *Journal of Electron Spectroscopy and Related Phenomena* **2005**, 144-147, 1179.

[15] A. Görling, *Physical Review A* **1999**, 59, 3359.

[16] a) G. Kresse, D. Joubert, *Physical review b* **1999**, 59, 1758; b) P. E. Blöchl, *Physical Review B* **1994**, 50, 17953.

[17] J. Tersoff, D. R. Hamann, *Physical Review B* **1985**, 31, 805.

[18] C. Huang, L. Xie, H. Zhang, H. Wang, J. Hu, Z. Liang, Z. Jiang, F. Song, *Nanomaterials*, 10.3390/nano12152518

[19] J. Lin, S. T. Pantelides, W. Zhou, *ACS Nano* **2015**, 9, 5189.

[20] F. Arnold, R.-M. Stan, S. K. Mahatha, H. E. Lund, D. Curcio, M. Dendzik, H. Bana, E. Travaglia, L. Bignardi, P. Lacovig, D. Lizzit, Z. Li, M. Bianchi, J. A. Miwa, M. Bremholm, S. Lizzit, P. Hofmann, C. E. Sanders, *2D Materials* **2018**, 5, 045009.

[21] a)J. Feng, D. Biswas, A. Rajan, M. D. Watson, F. Mazzola, O. J. Clark, K. Underwood, I. Marković, M. McLaren, A. Hunter, D. M. Burn, L. B. Duffy, S. Barua, G. Balakrishnan, F. Bertran, P. Le Fèvre, T. K. Kim, G. van der Laan, T. Hesjedal, P. Wahl, P. D. C. King, *Nano Letters* **2018**, 18, 4493; b)Y. Wang, J. Ren, J. Li, Y. Wang, H. Peng, P. Yu, W. Duan, S. Zhou, *Physical Review B* **2019**, 100, 241404.
[22] Y.-C. Lin, D. O. Dumcenco, Y.-S. Huang, K. Suenaga, *Nature Nanotechnology* **2014**, 9, 391.
[23] a)D. Li, X. Wang, C.-m. Kan, D. He, Z. Li, Q. Hao, H. Zhao, C. Wu, C. Jin, X. Cui, *ACS Applied Materials & Interfaces* **2020**, 12, 25143; b)Y. Wang, Z. Sofer, J. Luxa, M. Pumera, *Advanced Materials Interfaces* **2016**, 3, 1600433; c)G. V. Pushkarev, V. G. Mazurenko, V. V. Mazurenko, D. W. Boukhvalov, *Physical Chemistry Chemical Physics* **2019**, 21, 22647.
[24] A. Zangwill, *Physics at Surfaces*, Cambridge University Press, Cambridge **1988**.
[25] J. Cai, W. Jolie, C. C. Silva, M. Petrović, C. Schlueter, T. Michely, M. Kralj, T.-L. Lee, C. Busse, *Physical Review B* **2018**, 98, 195443.

# Phase Transition of single-layer vanadium diselenide on Au(111) with distinguished electronic structures

*Jinbang Hu, Xiansi Wang, Chaoqin Huang, Fei Song and Justin W Wells*

Jinbang Hu

Department of Physics, Norwegian University of Science and Technology, NO-7491 Trondheim, Norway

E-mail: jinbang.hu@ntnu.no

Xiansi Wang

Hunan University, Changsha, 410082, China

Fei Song, Chaoqin Huang

Shanghai Advanced Research Institute, Chinese Academy of Sciences, Shanghai 201000, China;

Justin W. Wells

Department of Physics, Norwegian University of Science and Technology, NO-7491 Trondheim, Norway

Semiconductor Physics, Department of Physics, University of Oslo (UiO), NO-0371 Oslo, Norway

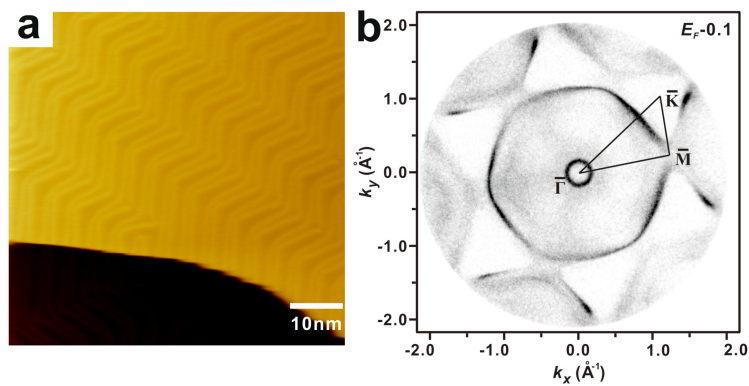

Figure S1. (a) STM image of clean Au(111) surface. (b) ARPES data of constant energy contours throughout the surface Brillouin zone at 0.1eV below the Fermi level.

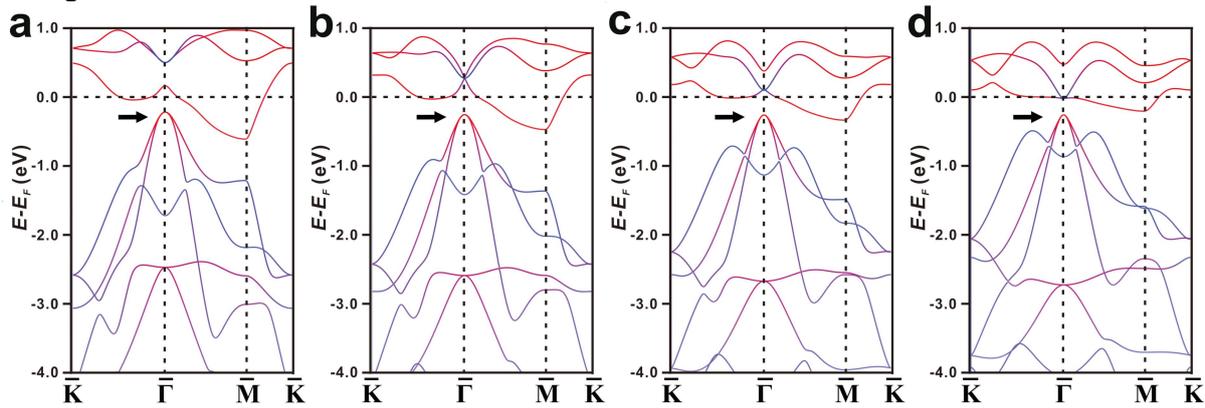

Figure S2. Theoretically calculated band structure of SL freestanding VSe$_2$ for expanded in-plane lattice parameters (a) 3.35Å, (b) 3.55Å, (c) 3.65Å, (d) 3.75Å. The relative energy position of the parabolically dispersing band around the Γ point is marked by the black arrow.

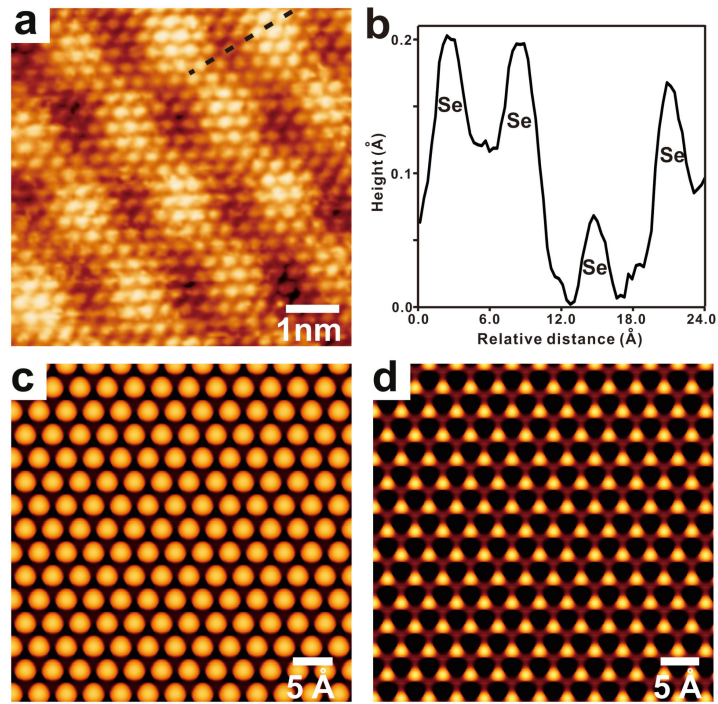

Figure S3. (a) Atomic-resolution STM image of SL VSe$_2$ showing the hexagonal lattice of Se atoms in the topmost sublayer of the VSe$_2$ sandwich-type structure. (b) The intensity profile along the black dashed line marked in (a). (c) Simulation of freestanding 1T VSe$_2$ and (d) 2H VSe$_2$. Scanning parameters: (a) −1.2 V, 0.12 nA.